\documentclass[submit]{smj}

\usepackage[english]{babel}
\usepackage{sidecap}
\usepackage[utf8]{inputenc}
\usepackage{amsfonts}
\usepackage{subfigure}
\usepackage{graphicx}
\usepackage{colortbl}
\usepackage{xcolor}
\usepackage{placeins}
\usepackage{url}
\usepackage{multicol}
\usepackage{multirow}
\usepackage[english,ruled,vlined]{algorithm2e}

\providecommand{\SetAlgoLined}{\SetLine}

\newcommand{\bx}{\mathbf{x}}

\newcommand{\bW}{\mathbf{W}}

\newcommand{\bZ}{\mathbf{Z}}

\newcommand{\bX}{\mathbf{X}}
\newcommand{\bB}{\mathbf{B}}
\newcommand{\bU}{\mathbf{U}}
\newcommand{\bV}{\mathbf{V}}

\newcommand{\bn}{\mathbf{n}}

\newcommand{\balpha}{\boldsymbol{\alpha}}
\newcommand{\bpi}{\boldsymbol{\pi}}
\newcommand{\bPi}{\boldsymbol{\Pi}}

\usepackage{dsfont}
\newcommand{\indi}{\mathds{1}}

\Author{Etienne C\^{o}me\Affil{1} 
        and Pierre Latouche\Affil{2} 
}
\AuthorRunning{Etienne C\^{o}me and Pierre Latouche}

\Affiliations{

\item Universit\'{e} Paris-Est, 
  IFSTTAR,    GRETTIA,  
  Noisy-Le-Grand, France

\item Laboratoire SAMM, 
      Universit\'e Paris 1 Panth\'{e}on-Sorbonne,
      France

}   

\CorrAddress{Pierre Latouche, 
             Laboratoire SAMM, 
             Universit\'e Paris 1 Panth\'{e}on-Sorbonne,
             90 rue de Tolbiac, F-75634 Paris Cedex 13, 
             France
             }
\CorrEmail{pierre.latouche@univ-paris1.fr}
\CorrPhone{(+33)\;1\;44\;07\;88\;26}
\CorrFax{(+33)\;1\;44\;07\;89\;35}

\Title{Model selection and clustering in stochastic block models with the exact integrated complete data likelihood}
\TitleRunning{Greedy inference in stochastic block models}

\Abstract{
The stochastic block model (SBM) is a mixture
model used for the clustering of nodes in networks. It has now been
employed for more than a decade to analyze very different types
of networks in many scientific fields such as Biology and social sciences.  Recently, a analytical expression, based
on the collapsing of the SBM model parameters, was proposed along with
a sampling  procedure which allows  the clustering of the  vertices as
well  as the  estimation of  the number  of clusters  to  be performed
simultaneously.  However,
while the corresponding algorithm can technically deal with up to ten
thousand nodes and millions of edges, the Markov chain tends to have poor mixing properties,
\emph{i.e.} low acceptance rates, for large networks. Therefore, the
number of clusters tends to be highly overestimated, even for a very
large number of samples.  In this paper, we rely on a similar expression, that we call the integrated complete data log likelihood,
and  we  propose  a   greedy  inference  algorithm  which  focuses  in
maximizing this exact quantity. The algorithm  has a better computational cost than existing
inference techniques for SBM and can be employed to analyze large
networks (several tens of thousands of nodes and million of edges) with no convergence issues.  Using toy data sets, the algorithm is shown to
improve  over existing  strategies, both  in terms  of  clustering and
model  selection. An  application on  a  network of  blogs related  to
illustrations and comics is also provided.
}


\begin{document}

\maketitle


\section{Introduction}

\subsection{Context}

There is a long history of research on networks which goes back to
the earlier work of \cite{bookmoreno1934}. Because they are simple
data structures yet capable of representing complex systems, they
are used in many scientific fields \citep{articlebarabasi2004,articlepalla2007}.
Originally considered in social sciences \citep{articlefienberg1981}
to characterize relationships between actors \citep{holland83,BouletEtAl2008Neurocomputing},
networks are also used to describe neural networks \citep{articlewhite1986},
powergrids \citep{articlewatts1998}, and the Internet \citep{adamic05,zanghi08a}.
Other examples of real networks can be found in Biology with the use
of regulatory networks to describe the regulation of genes by transcriptional
factors \citep{articlemilo2002} or metabolic networks to represent
pathways of biochemical reactions \citep{articlelacroix2006}. As
the number of networks used in practice has been increasing, a lot
of attention has been paid on developing graph clustering algorithms
to extract knowledge from their topology. Existing methods usually
aim at uncovering very specific patterns in the data, namely communities
or disassortative mixing. For an exhaustive review, we refer to \cite{articlegoldenberg2009}.

Most graph clustering algorithms look for communities, where two nodes
of the same community are more likely to be connected than nodes of
different communities. These techniques \citep{articlenewman2004,proceedingsnewman2006}
often maximize the modularity score proposed by  \cite{proceedingsgirvan2002}
for clustering. However, recent work of \cite{proceedingsbickel2009}
showed that they were asymptotically biased and tended to lead to
the discovery of an incorrect community structure, even for large
graphs. Alternative strategies, see for instance \cite{articlekrivitsky2009},
are generally related to  the probabilistic model of \cite{articlehandcock2007}
which generalizes the work of  \cite{articlehoff2002}. Nodes are first mapped
into a a latent space and then clustered depending on their latent
positions. Community structure algorithms are commonly used for affiliation
network analysis. As mentioned in \cite{newman07}, other graph clustering
algorithms aim at uncovering dissasortative mixing in networks where,
contrary to community structure, nodes mostly connect to nodes of
different clusters. They are particularly suitable for the analysis
of bipartite or quasi bipartite networks \citep{articleestrada2005}.

In this paper, we consider the stochastic block model (SBM) proposed
by  \cite{nowicki01} which is a probabilistic generalization \citep{articlefienberg1981,holland83}
of the work of \cite{articlewhite1976}. As
pointed out by \cite{daudin08},
SBM can be seen as a mixture model for graphs. It assumes that nodes
are spread into $K$ clusters and uses a $K\times K$ matrix $\mathbf{\Pi}$
to describe the connection probabilities between pairs of nodes. No
assumption is made on $\mathbf{\Pi}$ such that very different structures
can be taken into account. In particular, as shown in \cite{inbooklatouche2009},
contrary to the methods mentioned previously, SBM can be used to retrieve
both communities and disassortative mixing in networks.

\subsection{Prior work: models}

Many extensions have been developed to overcome some limits of the
standard SBM. For example, \cite{articlemariadassou2010} introduced
recently a probabilistic framework to deal with valued edges, allowing
covariates to be taken into account. While the first model they proposed
explains the value of an edge, between a pair of nodes, through
their clusters only, the second and third approaches do account for
covariates through Poisson regression models. This framework is relevant
in practice because extra information on the edges is sometimes available,
such as phylogenetic distances in host-parasite networks or amounts
of energy transported between nodes in powergrids. 

Another drawback
of SBM is that it assumes that each node belongs to a single cluster
while many objects in real world applications belong to several groups
or communities \citep{articlelatouche2011}. To tackle this issue
 \cite{airoldi08} proposed the mixed membership stochastic block
model (MMSBM) \citep{proceedingsairoldi2006,articleairoldi2007}.
A latent variable $ $$\bpi_{i}$, drawn from a Dirichlet distribution,
is associated to each node $i$ of a network. Given a pair $(i,j)$
of nodes, two binary latent vectors $\bZ_{i\rightarrow j}$ and $\bZ_{i\leftarrow j}$
are then considered. The vector $\bZ_{i\rightarrow j}$ is assumed
to be sampled from a multinomial distribution with parameters $(1,\mathbf{\bpi_{i})}$
and describes the cluster membership of $i$ in its relation towards
$j$. By symmetry, $\bZ_{i\leftarrow j}$ is drawn from multinomial
distribution with parameters $(1,\mathbf{\bpi_{j})}$ and characterizes
the cluster membership of $j$ in its relation towards $i$. Thus, in
MMSBM, since each node can have different latent vectors through its
relations towards other nodes, it can belong to several clusters. The
connection probability between $i$ and $j$ is finally given by $p_{ij}=\bZ_{i\rightarrow j}^{\intercal}\bB\bZ_{i\leftarrow j}$, with $\bB$ a matrix of parameters to estimate.
The  overlapping  stochastic  block  model  (OSBM)  was  proposed  by
\cite{articlelatouche2011}
as an alternative probabilistic model for networks allowing overlapping
clusters. Contrary to MMSBM, edges are influenced by the fact that
some nodes belong to multiple clusters. Thus, each node $i$ is characterized
by a binary latent vector $\bZ_{i}$ sampled from a product of Bernoulli
distributions. An edge between $ $nodes $i$ and $j$ is then drawn
from a Bernoulli distribution with parameter $p_{ij}=g(a_{\bZ_{i},\bZ_{j}})$.
The function $g(\cdot)$ is the logistic function while $a_{\bZ_{i},\bZ_{j}}$
is a real variable describing the interactions between the nodes,
depending on the different clusters they are associated with.  It is given by
$a_{\bZ_{i},\bZ_{j}}=\bZ_{i}^{\intercal}\bW\bZ_{j}+\bZ_{i}^{\intercal}\bU+\bV^{\intercal}\bZ_{j}+W^{*}$, where $\bW,\bU,\bV$ and $W^{*}$ are matrices of parameters. Finally,
we mention the work of \cite{articlekarrer11} who proposed an interesting extension of SBM to deal with node degree heterogeneity inside clusters. 
The  model deals with valued edges and includes another set of parameters describing vertices attractiveness. Using the right constraints the model is identifiable (up to permutations of clusters) and the connectivity parameters can be directly related to vertices degree. This work was extended to oriented networks in \cite{zhu2012} and finally tools for model selection between different models are derived in \cite{yan2012}.

\subsection{Prior work: inference}

While  research has focused  in the  last few  years on  proposing new
types of  SBM models, \cite{mcdaid11}  chose to consider  the standard
SBM model  and to focus  on the inference  task. They proposed  a new
inference procedure, that we shall discuss shortly in Section \ref{ssec:contrib}, for which very encouraging results have been
obtained. Following their work, we consider in this paper the standard
SBM model, which has been widely used in practice for network analysis,
for more than a decade. We aim at developing a new optimization
procedure,   improving  over   existing  inference   strategies.  This
framework can be extended to other types of SBM models.

In SBM, the
posterior distribution over the latent variables, given the parameters
and the observed data, cannot be factorized due to conditional dependency.
Therefore, optimization techniques such as the expectation maximization
(EM) algorithm cannot be used directly for clustering. To tackle this
issue, \cite{daudin08} proposed an approximation method based on
a variational EM algorithm. Note that an on line version of this algorithm
exists \citep{zanghi08b}. A Bayesian framework was also considered
by  \cite{nowicki01} where conjugate priors for the model parameters
were introduced. Again, because the posterior distribution over the
model parameters, given the data, is not tractable, approximation
techniques were employed for inference. Thus,  \cite{nowicki01} used
a Gibbs sampling procedure while \cite{articlelatouche2012} relied
on a variational Bayes EM  algorithm. Note that a similar approach was
considered by \cite{hofman08} for a constrained SBM model. Two model
selection criteria, the integrated classification likelihood (ICL) and the integrated likelihood variational Bayes (ILvb) have been developed for
SBM   in  order   to  estimate   the   number  $K$   of  clusters   in
networks.   Alternative   strategies   using   spectral   partitioning
\citep{fishkind2012} or hypothesis testing \citep{bickel2013} have also been considered recently. Standard
criteria such as the Akaike information criterion (AIC) or Bayesian
information criterion (BIC) cannot be used because they rely on the
SBM observed data log likelihood which is not tractable in practice (see for instance
\cite{inbooklatouche2009}). ICL was originally
developed by \cite{articlebiernacki2000} for Gaussian mixture models
and then adapted by \cite{daudin08} to SBM. It is based on Laplace
and Stirling
approximations of the integrated complete data log likelihood. As shown in \cite{articlebiernacki2010},
it tends to miss some important structures present in the data for
small data sample, because of the asymptotic approximations.
To tackle this drawback, \cite{articlelatouche2012} proposed the
ILvb criterion which relies on a variational Bayes approximation of
the integrated observed data log likelihood. 

\subsection{Contributions}\label{ssec:contrib}

An alternative  inference strategy was  proposed recently for  the SBM
model  in \cite{mcdaid11}.  The  authors first  derived an  analytical
expression, based on the collapsing of the SBM model parameters. Then,
they relied on an allocation sampler algorithm as in \cite{nobile2007}
which allows the clustering of  the vertices as well as the estimation
of  the  number  of  clusters  to be  performed  simultaneously.  This
sampling procedure was shown to improve over existing inference strategies for the SBM
model, both in terms of clustering and model selection. However, while
the   algorithm can  technically  deal  with  up to  ten
thousand nodes  and millions of edges, the  corresponding Markov chain
tends  to  have poor  mixing  properties,  \emph{i.e.} low  acceptance
rates,  for such  large networks.  In  practice, some  moves aiming  at
reducing the model complexity  are rarely accepted, the convergence of
the chain  is slow and  therefore the number  of clusters tends  to be
highly overestimated, even for a very large number of samples. In this
paper, we aim at avoiding this issue. We rely on a similar analytical expression, that we
call the  integrated complete  data log likelihood.  The corresponding
criterion is denoted $ICL_{ex}$ where $ex$ stands for exact. Contrary to the ICL criterion in
\cite{daudin08}, $ICL_{ex}$ does not rely on any asymptotic approximations.
 We then propose
a greedy inference algorithm which maximizes this exact quantity.

Contrary   to   the clustering   algorithms   of    \cite{daudin08}   and
\cite{articlelatouche2012}, we emphasize that the algorithm we propose
maximizes  an analytical  criterion and  does  not rely  on any  lower
bounds for approximation. We recall that the lower bound of the variational EM algorithm
proposed by  \cite{daudin08} approximates the observed data log likelihood,
while \cite{articlelatouche2012} introduced a lower bound to estimate
the  integrated observed  data log  likelihood.  Moreover,  the greedy
search has the advantage of  performing the clustering of the vertices
and the estimation  of the number of clusters at the  same time, as in
\cite{mcdaid11}, and no model
selection criterion has to be computed for various values of $K$.
Starting from a complex model with $K=K_{up}$ clusters, ($K_{up}$
being an upper bound for $K$), the proposed algorithm swaps labels
until $ICL_{ex}$ reaches a local maximum. During the process, clusters
may disappear, \textit{i.e.} their cardinality reaches zero. Such an approach leads
to a simple and time attractive algorithm with complexity of $\mathcal{O}(L+NK_{up}^{2})$,
with $L$ the total number
of edges in the network and $N$ the number of vertices. Thus, it has a better computational cost than existing
inference  techniques for  SBM and  can be  employed to  analyze large
networks,  while not  having  to deal  with  the convergence  sampling
issues of \cite{mcdaid11}.


As we shall see through a series of experiments, the greedy algorithm
takes benefit of computing the exact ICL and improves over existing
methods, both in terms of clustering and model selection. It can also
deal  with large  networks  with  tens of  thousands  of vertices  and
millions of edges.

\section{The stochastic block model }

We consider a binary network with $N$ nodes represented by its adjacency
matrix $\bX$ such that $X_{ij}=1$ if there is an edge from node
$i$ to node $j$, 0 otherwise. In this paper, we focus on directed
networks, \textit{i.e.} relations are oriented. Therefore $\bX$ is not symmetric.
Moreover, we do not consider any self loop, that is an edge from a
node to itself. We emphasize that all the optimization equations derived
in this work can easily be adapted to deal with undirected networks
or to take into account self loops.

\subsection{Model and notations}

The stochastic block model (SBM) introduced by  \cite{nowicki01}
assumes that the nodes are spread into $K$ clusters with connectivity
patterns described by a $K\times K$ matrix $\bPi$. The cluster of each
node is given by its binary membership vector $\bZ_{i}$ sampled from
a multinomial distribution :
\[
\bZ_{i}\sim\mathcal{M}(1,\balpha=(\alpha_{1},\dots,\alpha_{K})),\sum_{k=1}^{K}\alpha_{k}=1,
\]
such that $Z_{ik}=1$ if $i$ belongs to cluster $k$ and zero otherwise.
Contrary to the work of \cite{articlelatouche2011}, each node belongs to a single
cluster, that is $\sum_{k=1}^{K}Z_{ik}=1,\forall i$. Given the vectors
$\bZ_{i}$ and $\bZ_{j}$, an edge between node $i$ and $j$ is then
drawn from a Bernoulli distribution with probability $\Pi_{kl}$ :
\[
X_{ij}|Z_{ik}Z_{jl}=1\sim B(\Pi_{kl}).
\]
This leads to a simple yet flexible generative model for networks.
First, all the vectors $\bZ_{i}$ are sampled independently. We denote by
$\bZ$ the binary $N\times K$ matrix storing the $\bZ_{i}$s as
raw vectors : 

\begin{equation}\label{eq:pZalpha}
p(\bZ|\balpha)= \prod_{i=1}^{N}\mathcal{M}(\bZ_{i};1,\balpha) =  \prod_{i=1}^{N}\prod_{k=1}^{K}\alpha_{k}^{Z_{ik}}.
\end{equation}
Then, given the latent structure $\bZ$, all the edges in $\bX$ are drawn independently: 
\begin{equation}\label{eq:pXpi}
\begin{aligned}
p(\bX|\bZ,\bPi)= & \prod_{i\neq j}^{N}p(X_{ij}|\bZ_{i},\bZ_{j},\bPi)\\
= & \prod_{i\neq j}^{N}\prod_{k,l}^{K}\mathcal{B}(X_{ij};\Pi_{kl})^{Z_{ik}Z_{jl}}\\
= & \prod_{i\neq j}^{N}\prod_{k,l}^{K}\left(\Pi_{kl}^{X_{ij}}(1-\Pi_{kl})^{1-X_{ij}}\right)^{Z_{ik}Z_{jl}}.
\end{aligned}
\end{equation}

\subsection{Integrated classification likelihood criteria}


In this paper, we will consider the integrated complete data log likelihood
$\log p(\bX,\bZ|K)$
in order to focus on the inference of $\bZ$ and $K$ from the observed
data $\bX$, all the  SBM parameters $(\balpha, \bPi)$ being integrated
out.    We   point  out   that   a   similar   quantity  appeared   in
\cite{mcdaid11}, in a different context and for different purposes, when the authors derived posterior distributions of
an allocation sampler algorithm, as is \cite{nobile2007}. 

We first recall existing
approximations and then derive in Section \ref{ssec:exactICL}  the integrated complete data log likelihood.

\subsubsection{Asymptotic ICL criterion}

When considering a factorized prior distribution $p(\balpha,\bPi|K)=p(\balpha|K)p(\bPi|K)$ over the
model  parameters, as  in \cite{articlebiernacki2000},  the integrated
complete data log likelihood easily decomposes into two terms:
\begin{equation}\label{eq:integratedComp}
  \begin{aligned}
\log p(\bX,\bZ|K)&=\log \left(\int_{\balpha,\bPi} p(\bX,\bZ,\bPi,\balpha|K)d\balpha
d\bPi\right) \\
&=                   \log\left(            \int_{\bPi}p(\bX|\bZ,\bPi,K)p(\bPi|K)d\bPi
\int_{\balpha}p(\bZ|\balpha,K)p(\balpha|K)d\balpha \right)\\
&= \log p(\bX|\bZ,K) + \log p(\bZ|K).
\end{aligned}
\end{equation}
However,  for an  arbitrary choice  of the  priors  $p(\balpha|K)$ and
$p(\bPi|K)$,  the  marginal distributions  $p(\bX|\bZ,K)$  as well  as
$p(\bZ|K)$ are usually not tractable and (\ref{eq:integratedComp}) does not
have any analytical form.  To tackle this issue,
 \cite{daudin08} relied on an asymptotic approximation of $\log p(\bX, \bZ|K)$, so called integrated classification
likelihood criterion (ICL).  Note that ICL was originally proposed by
 \cite{articlebiernacki2000} for Gaussian mixture models.  It was then
adapted by  \cite{articlebiernacki2010} to mixtures of multivariate
multinomial distributions and  to the SBM model by  \cite{daudin08}. In the case we consider of a directed graph without
self-loop, ICL is given by:
\begin{equation}\label{eq:iclapprox}
  \begin{aligned}
    ICL(\bZ, K)& \approx \log p(\bX,\bZ|K) \\
    &=      \max_{\balpha,\bPi}     \log     p(\bX,\bZ|\balpha,\bPi,K)
    -\frac{1}{2}K^{2}\log \left(N(N-1)\right) - \frac{K-1}{2}\log(N).
  \end{aligned}
\end{equation}
 For  an extensive  description of  the  use of  Laplace and  Stirling
 approximations   to   derive  the   ICL   criterion,   we  refer   to
 \cite{articlebiernacki2000}.  Since it approximates  the integrated
 complete  data  log  likelihood,  ICL  is known  to  be  particularly
 suitable when  the focus  is on  the clustering task  and not  on the
 estimation   of   the   data    density.   However,   as   shown   in
 \cite{articlebiernacki2010,articlemariadassou2010},   it  tends   to  miss   some  important
 structures present in the data because of the (asymptotic) approximations.

 We emphasize that ICL is only used in the literature as a
model selection criterion. In practice, a clustering method such as an EM like algorithm
 for instance is employed to obtained some estimates $\tilde{\bZ}$ of $\bZ$, for various
 values of the  number $K$ of classes. ICL is  then computed for every
 pair $(\tilde{\bZ},K)$ and the pair $(\tilde{\bZ}^{*},K^{*})$ is chosen such
 that the criterion is maximized.  Thus, ICL is optimized only through
 the   results   $(\tilde{\bZ},K)$    produced   by   the   clustering
 algorithm. Conversely, after having given an analytical expression $ICL_{ex}$ of
 the integrated complete  data log likelihood in the  next section, we
 will show in Section \ref{sec:greedy} how to optimize directly $ICL_{ex}$
 with  respect  to  $\bZ$  and  $K$.  As shown  in  Section  \ref{ssec:complexity},  this
 significantly reduces the computational cost of the inference procedure.

\subsubsection{Exact ICL criterion}\label{ssec:exactICL}

We rely on the same Bayesian framework as in \cite{nowicki01} and
\cite{inbooklatouche2009}. Thus, we consider non informative conjugate
priors for the model parameters $\balpha$ and $\bPi$. Since $\balpha$,
describing  the   cluster  proportions,  parametrizes   a  multinomial
distribution  (\ref{eq:pZalpha}),   we  rely  on   a  Dirichlet  prior
distribution:
\begin{equation*}
  p(\balpha) = \mathrm{Dir}\left(\balpha;\: \bn^{0}=(n_{1}^{0},\dots,n_{K}^{0})\right).
\end{equation*}
A   common  choice    consists   in   fixing  the
hyperparameters to $1/2$,  \emph{i.e.} $n_{k}^{0}=1/2,\forall k$. Such
a distribution  corresponds to a non informative  Jeffreys prior which
is  known  to  be  proper \citep{proceedingsjeffreys1946}.  A  uniform
distribution can  also be obtained  by setting the  hyperparameters to
$1$.

Moreover, since  the presence or absence  of an edge  between nodes is
sampled from  a Bernoulli  distribution, we consider  independent Beta
prior distributions to model the connectivity matrix $\bPi$:
\begin{equation*}
  p(\bPi)=\prod_{k,l}^{K}\mathrm{Beta}(\Pi_{kl};\:\eta_{kl}^{0},\zeta_{kl}^{0}).
\end{equation*}
Again,  if  no prior  information  is  available, all  hyperparameters
$\eta_{kl}^{0}$  and $\zeta_{kl}^{0}$  can be  set  to $1/2$  or 1  to
obtain a Jeffreys or uniform distribution.

With  these choices of  conjugate prior  distributions over  the model
parameters,  the  marginal   distributions  $p(\bX|\bZ,K)$  as  well  as
$p(\bZ|K)$ in (\ref{eq:integratedComp}) have analytical forms, and so
has the integrated complete data  log likelihood, as proved in \ref{ann:posterior}. We   call   $ICL_{ex}$  the
corresponding criterion, where $ex$ stands for exact. It is given by:
\begin{equation}\label{eq:exICL}
  \begin{aligned}
    ICL_{ex}(\bZ,K) &= \log p(\bX,\bZ|K)\\
    &=
    \sum_{k,l}^{K}\log\left(\frac{\Gamma(\eta_{kl}^{0}+\zeta_{kl}^{0})\Gamma(\eta_{kl})\Gamma(\zeta_{kl})}{\Gamma(\eta_{kl}+\zeta_{kl})\Gamma(\eta_{kl}^{0})\Gamma(\zeta_{kl}^{0})}\right)
    + \log\left(\frac{\Gamma(\sum_{k=1}^{K}n_{k}^{0})\prod_{k=1}^{K}\Gamma(n_{k})}{\Gamma(\sum_{k=1}^{K}n_{k})\prod_{k=1}^{K}\Gamma(n_{k}^{0})}\right),
  \end{aligned}
\end{equation}
 where the components $n_{k}$ are:
\begin{equation*}
n_{k}=n_{k}^{0}+\sum_{i=1}^{N}Z_{ik}, \forall k \in \{1,\dots,K\},
\end{equation*}
and can  be seen  as pseudo counters  of the  number of nodes  in each
class. Moreover, the parameters
$(\eta_{kl},\zeta_{kl})$ are given by:
\begin{equation*}
\eta_{kl}=\eta_{kl}^{0} + \sum_{i\neq
  j}^{N}Z_{ik}Z_{jl}X_{ij}, \forall (k,l) \in \{1,\dots,K\}^{2},
\end{equation*}
and
\begin{equation*}
\zeta_{kl}=\zeta_{kl}^{0}    +
\sum_{i\neq   j}^{N}  Z_{ik}Z_{jl}(1-X_{ij}),   \forall  (k,l)  \in
\{1,\dots,K\}^{2}.
\end{equation*}
They represent  pseudo counters  of the number  of edges  and non-edges
connecting nodes of class $k$ to nodes of class $l$, respectively. 

Since  $ICL_{ex}(\bZ,K)$  involves a  marginalization  over the  model
parameters $\balpha$ and $\bPi$, which have non informative priors, it
automatically  penalizes  the  number  $K$ of  classes  and  therefore
controls   the   model   complexity.   Indeed,   as   highlighted   by
\cite{articlebiernacki2000} in  the case of  standard Gaussian mixture
models, the  penalization terms  are encompassed through  the use  of the
gamma  function.  For example, replacing the gamma function
$\Gamma(\cdot)$   with   the   Stirling  approximation  $\Gamma(t+1)\approx
t^{t+1/2}\exp(-t)(2\pi)^{1/2}$,     in    the    second     term    in
(\ref{eq:exICL}),  would reveal  the  penalization $(1/2)(K-1)\log
N$ in  (\ref{eq:iclapprox}).  Similarly,  replacing the first  term in
(\ref{eq:exICL}) using  such an asymptotic  approximation would reveal
the penalization $(1/2)K^{2}\log(N(N-1))$ in (\ref{eq:iclapprox}). For more details,
we refer to \cite{articlebiernacki2000}.

Note that maximizing  $ICL_{ex}(\bZ,K)=\log p(\bX,\bZ|K)$ with respect
to $\bZ$ only is
equivalent to maximizing $\log p(\bZ|\bX,K)$ since $\log p(\bX,\bZ|K)=\log
p(\bZ|\bX,K) +  \log p(\bX|K)$. However, while  $\log (\bX,\bZ|K)$ has
an analytical form, $\log p(\bZ|\bX,K)$ has not.  Therefore, existing
algorithms for SBM,  relying on $p(\bZ|\bX,K)$, have had to consider approximation techniques like
Gibbs sampling or variational bounds, for inference purposes, while we
consider here an exact quantity. Moreover, we point out  that the $ICL_{ex}$ criterion is related to the variational Bayes
approximation  of the  integrated observed  data log  likelihood $\log
p(\bX|K)$ proposed by \cite{articlelatouche2012}. The key difference
is that the parameters $(n_{k},\eta_{kl},\zeta_{kl})$ in
$ICL_{ex}$ depend
on  the hard  assignment $\bZ$  of  nodes  to classes  and not  on
approximated  posterior  probabilities  $\boldsymbol{\tau}$. Moreover,  $ICL_{ex}$
does      not     involve     any      entropy     term      as     in
\cite{articlelatouche2012}. 

\section{Greedy optimization} \label{sec:greedy}

Since the
model parameters have been marginalized out, the $ICL_{ex}$
criterion  only  involves the  cluster  indicator  matrix $\bZ$  whose
dimensionality depends on the number $K$ of clusters.  Thus, this
integrated likelihood is only a function of a partition $\mathcal{P}$, \emph{i.e.}
an assignment of
the vertices  to clusters.  Looking directly for  a global  maximum of
$ICL_{ex}$ is
not feasible  because it involves testing every  possible partition of
the vertices with
various values of $K$. However, this is a combinatorial problem for which
heuristics exist  to obtain  local maxima. In  this paper, we  rely on
greedy  heuristics which  have been  shown to  scale well  with sample
sizes  \citep{articlenewman2004}.  These approaches have already been used for graph
clustering using  \textit{ad-hoc} criteria such as modularity
\citep{articlenewman2004,blondel08} and are reminiscent of the well
known iterated conditional modes algorithm of \cite{besag86} used for
maximum \emph{a posteriori} estimation in Markov random fields. 

The algorithm (see Algorithm \ref{algo:greedy}) starts with a SBM model with $K=K_{up}$
clusters,   $K_{up}$  being   an  upper   bound  for   the   number  of
clusters. $K_{up}$ is assumed to be given as an input along with a $N \times K_{up}$
matrix $\bZ$. In practice, $K_{up}$ is set to a large value using user knowledge on the problem at hand, while $\bZ$ can be  initialized with the
methods described in the next section. The algorithm then cycles randomly through all the vertices of the
network.  At each step, a single node $i$ is considered while all the
membership vectors $\bZ_{j}$ for $j \neq i$ are hold fixed. If $i$ is currently in
cluster $g$, the method looks for every possible label swapping,
\emph{i.e.}  removing $i$ from cluster $g$ and assigning it to a cluster $h \neq g$, and computes the
corresponding change $\Delta_{g \rightarrow h}$ in the $ICL_{ex}$ criterion. Note
that $\Delta_{g \rightarrow h}$ takes two forms (see \ref{ann:swap}) whether cluster $g$ is empty after
removing $i$ or not. If no label swapping
induces  an increase  of the  criterion, the  vector $\bZ_{i}$  is not
modified. Otherwise,  the label swapping with the  maximal increase is
applied  and $\bZ_{i}$  is  changed accordingly.  During the  process,
clusters  may  disappear,   \textit{i.e.}  their  cardinality  reaches
zero. Each time  one of these moves is accepted,  the model is updated
and the corresponding column is removed from the cluster indicator matrix $\bZ$. 
 Finally, the algorithm stops if a complete pass over the vertices did
 not lead to any increase of the $ICL_{ex}$ criterion. Thus, the
algorithm, automatically infers the number of clusters while clustering
the vertices of the network. Starting  with an  over-segmented initial solution  our approach
simplifies  the model  until a  local  maximum is  reached.

\begin{algorithm}[h!]
\caption{Greedy ICL}
        \SetAlgoLined
        Set $K=K_{up}$ ; $\mathrm{stop}=0$ \;
        Initialize  the  $N  \times  K_{up}$ matrix  $\bZ$  ;  Compute
        $\boldsymbol{\eta}$,$\boldsymbol{\zeta}$,$\bn$ \;
	\While{$\mathrm{stop} \neq 1$}{
          $V=\{1,\dots,N\}$ ; $\mathrm{stop}=1$ \;
          \While{$V$ not empty}{
            Select a node $i$ randomly in $V$ ;  Remove $i$ from $V$ \;
            If $i$ is in cluster $g$, compute all terms $\Delta_{g \rightarrow h},\forall h \neq g$ \;
            \If{at least one $\Delta_{g \rightarrow h}$ is positive}{
               $\mathrm{stop}=0$ \;
              Find $h$ such that $\Delta_{g \rightarrow h}$ is maximum \;
              Swap labels of $i$: $Z_{ig}=0$ and $Z_{ih}=1$ \;
              \If{$g$ is empty}{
                Remove column $g$ in $\bZ$ ; Set $K=K-1$ \;
              }
              Update rows and columns $(g,h)$ of the matrices $\boldsymbol{\eta}$
              and $\boldsymbol{\zeta}$ \;
              Update the components $g$ and $h$ of vector $\bn$\;
}
         }
        }
\KwResult{$(\bZ,K)$}
\label{algo:greedy}
\end{algorithm}

We would like to highlight that a greedy algorithm was also considered
for     the     clustering     of     nodes     in     networks     by
\cite{articlekarrer11}. There are 
two  main differences  between their  approach and  ours,
apart from the models considered. First,  they assumed the number $K$ of classes to be known while we infer
it  through the  optimization process.  Moreover, replacing  the model
parameters in the observed-data likelihood, by their maximum likelihood
estimates, they considered a profile  likelihood, whereas we rely on an
integrated likelihood. As pointed  out in their conclusion, integrated
likelihoods are more suitable for model selection.

\subsection{Complexity}\label{ssec:complexity}
In order to set up such an algorithm, it is sufficient to know how to compute the changes in the $ICL_{ex}$ criterion induced by the possible swap movements (from cluster $g$ to cluster $h$) for a given node $i$, the others being kept fixed. Such changes can be computed efficiently (see \ref{ann:swap} for details) and the complexity of finding the best swap movement for a node is in average $\mathcal{O}(l+K^2)$, where $l$ is the average number of edges per node. Such complexity can be achieved, since good approximations of the logarithm of the gamma function are available with constant running time. The greedy  algorithm   has   therefore  a total  complexity   of $\mathcal{O}(N(l  +K_{up}^2)+L)$,  since   a  swap  movement  cost  is
$\mathcal{O}(l+K^2)$;  the   initialization  of  the   edges  counters
$(\eta_{kl},\zeta_{kl})$ cost  is $L$ (the  total number of edges  in the
graph) and several complete passes over the set of nodes will be performed
(typically  less  than 10).  Eventually,  this  can  be simplified  in
$\mathcal{O}(NK_{up}^2+L)$, since $K_{up}^2$ may certainly dominate $l$ and  compared   to   the  complexity   of
$\mathcal{O}(LK_{up}^3)$ achieved using  a variational algorithm and a
model           selection           criterion          as           in
\cite{daudin08,articlelatouche2012}. Indeed,  contrary to our approach
which  estimates  the  number  of  clusters in  a  single  run,  while
clustering  the nodes,  these approaches  are run  multiple  times for
various  values   of  $K$  and   $K^{*}$  is  chosen  such   that  the
corresponding model  selection criterion is maximized.  Since each run
costs     $\mathcal{O}(LK^2)$,    the     overall     complexity    is
$\mathcal{O}(LK_{up}^3)$.

\subsection{Initialization and restarts}
 Several solutions are possible for initializing the algorithm, a
 simple choice consisting in sampling random partitions while a more
 relevant though  expensive starting point can be obtained with the k-means
 algorithm (using the adjacency matrix by rows as input and a classical euclidean distance). One possible trade-off in terms of computational burden is
 to use only  few iterations of k-means. We used  the latter method in
 all the experiments  that we carried out. Moreover,  since our method
 is only guaranteed  to reach a local optima, a  common strategy is to
 run the optimization algorithm with multiple initializations and to keep
 the best one according to the $ICL_{ex}$ criterion.

\subsection{Hierarchical clustering}
Eventually,  in a  final  step, it  is  possible to  check that  merge
movements between clusters do not induce any increase of the objective function. This can be done with a greedy
hierarchical algorithm which  costs $\mathcal{O}(K^3)$ (see details in
\ref{ann:fusion}). Since the labels swap algorithm usually 
greatly  reduces   the  number  of  clusters  ($K   <<  K_{up}$),  the
computational cost of this last
step is low.

\section{Experiments on synthetic data}

To  assess the  greedy  optimization method,  a  simulation study  was
performed and the proposed solution  was compared with available implementations of algorithms for SBM inference:
\begin{itemize}
\item  \textbf{vbmod}, \citep{hofman08}, a  variational-based approach
  dedicated to the search of community structures, implemented in Matlab and C. The
  random graph model they considered  can be seen as a constrained SBM
  where all  terms on the  diagonal of the connectivity  matrix $\bPi$
  are set to a unique parameter $\lambda$ and off-diagonal
  terms to another parameter $\epsilon$,
\item  \textbf{mixer}, \citep{daudin08}, another  variational approach
  but one which can deal with the standard SBM model (not only community structures) implemented in R and C, 
\item \textbf{colsbm}, \citep{mcdaid11},  a collapsed Gibbs sampler for SBM in
  C.  The  last version  of  the code  is  used  in this  experimental
  section.  It  involves  an  additional  move type  compared  to  the
  algorithm described in the associated publication. This move was found to greatly enhance the results.
\end{itemize}

 Our goal here  is to evaluate the
ability of  the different solutions to recover  a simulated clustering
\emph{without}  knowing the  number of  clusters. Only  a reasonable
upper bound $K_{up}$ on $K$ will be provided to the algorithms when needed. We recall that the
variational methods optimize  a lower bound for various  values of $K$
and select $K^*$  such that a model selection  criterion is maximized:
ICL for \textbf{mixer} and ILvB for \textbf{vbmod}. Conversely, the collapsed
Gibbs  sampler automatically provides  an estimate  of $K$,  since the
posterior of $K$ is made available. 

As a baseline, we also  compared our approach with a standard spectral
clustering approach \citep{Shi00}. Note that we supplied this spectral
approach with the true number of clusters, in all simulations.

 The  performances   are  assessed  in  terms   of  normalized  mutual
 information \citep[see][for details and justification of this measure
 to compare partitions]{vinh10} between the estimated cluster membership matrix $\bZ^{e}$ and the simulated one $\bZ^{s}$. The mutual information $I(\bZ^{e},\bZ^{s})$ between two partitions is to this end defined by:
\begin{equation}
I(\bZ^{e},\bZ^{s})=\sum^K_{k,l}p_{kl}\log\left(\frac{p_{kl}}{p^{e}_k p^{s}_l}\right),
\end{equation} 
with
\begin{equation}
p_{kl}=\frac{1}{N}\sum_{i,j}^NZ_{ik}^{e}Z_{jl}^{s},\,p^{e}_{k}=\frac{1}{N}\sum_{i=1}^NZ_{ik}^{e},\,p^{s}_{l}=\frac{1}{N}\sum_{i=1}^NZ_{il}^{s}.\nonumber
\end{equation}
The measure  $I(\bZ^{e},\bZ^{s})$ describes how  much is learnt  about the
true  partition  if  the  estimated  one is  known,  and  \textit{vice
  versa}. The  mutual information is  not an ideal  similarity measure
when the  two partitions  have a different  number of clusters and it is therefore preferable to use a normalized version of the mutual information such as:
\begin{equation}
NI(\bZ^{e},\bZ^{s})=\frac{I(\bZ^{e},\bZ^{s})}{\max\left(H(\bZ^{e}),H(\bZ^{s})\right)},
\end{equation} 
with  $H(\bZ)=-\sum_{k=1}^{K}  p_k  \log(p_k)$ and  $p_k=\frac{1}{N}\sum_i^N Z_{ik}$.  The
performances are evaluated on simulated clustering problems of varying
complexity and with different settings, in order to give insights about
the influence of the number $K$ of clusters, of the number of vertices
$N$ and of the type of connectivity  matrix $\bPi$. 

\subsection{Setting 1: small scale community structures}

\begin{figure}[h!]
\centering
\begin{tabular}{c}
\includegraphics[height=6.5cm]{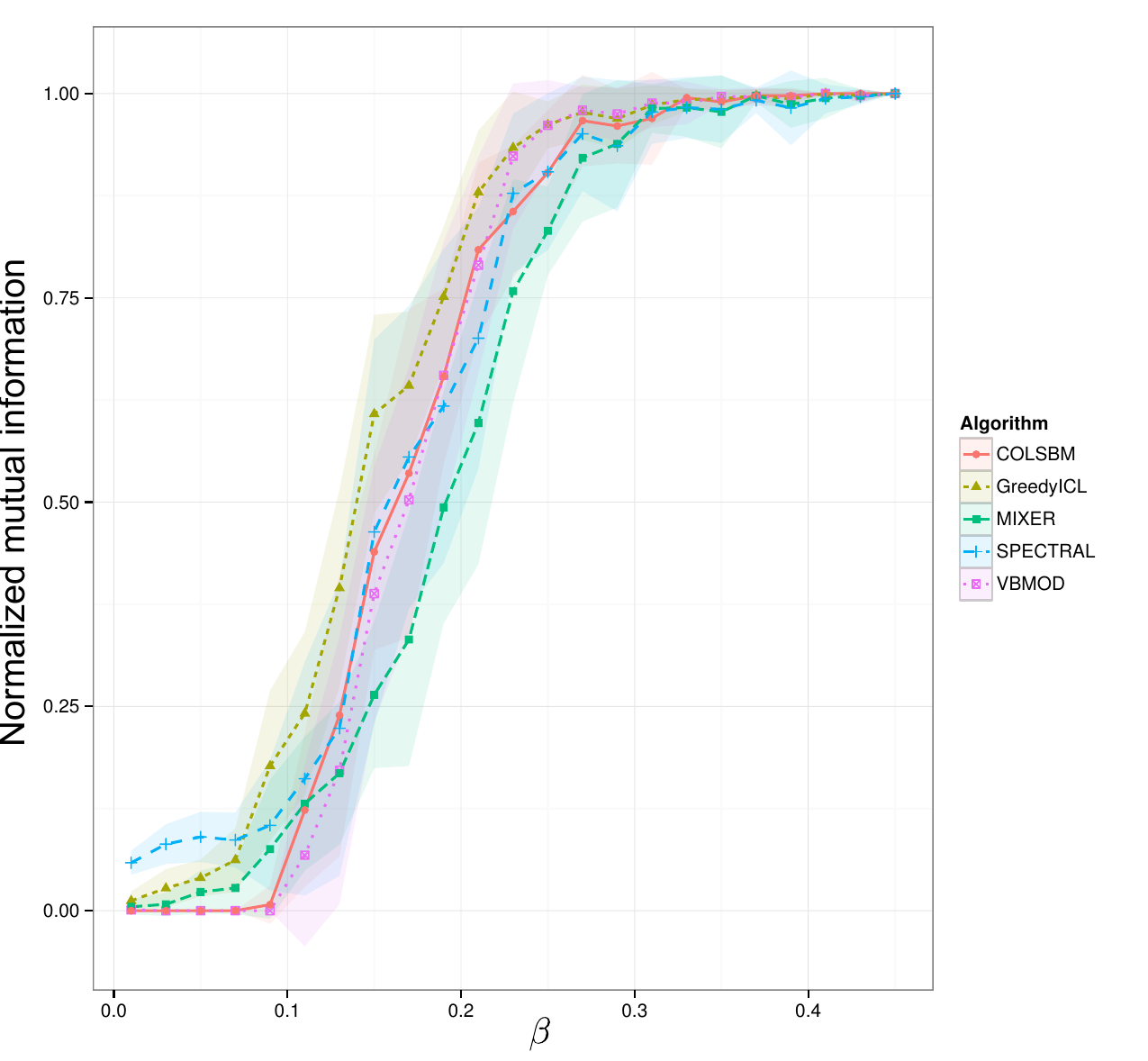}\\
Setting 1
\end{tabular}
\caption{\label{fig:simu1}Mean of mutual information between estimated
  and true cluster membership matrices using $20$ simulated graphs for
  each value of  $\beta$ in $\{0.45,0.43,\hdots,0.03,0.01\}$, and with
  $N=100,   K=5,   \epsilon=0.01$   for   the   different   algorithms
  \textbf{greedy    ICL},    \textbf{vbmod},    \textbf{colsbm}    and
  \textbf{mixer}.  The spectral  clustering approach  was run  with the
  true number of clusters, and used as a baseline.}
\end{figure}

The  first setting is  a classical  community simulation  with $N=100$
vertices  and  $K=5$ clusters.  The  cluster  proportions  are set  to
$\balpha=(1/5,1/5,1/5,1/5,1/5)$  and the  connectivity matrix  takes a
diagonal   form   with    off-diagonal   elements   equal   to   0.01:
$\mathbf{\Pi}_{kl}=0.01, \forall k \neq l$ and diagonal elements given
by $\mathbf{\Pi}_{kk}=\beta,\forall k$. $\beta$ is a complexity tuning
parameter  which ranges from  $0.45$ to  $0.01$. When  $\beta$ reaches
0.01,  the  model is  not  identifiable  (the  connectivity matrix  is
constant) and the true cluster memberships cannot be recovered. The set
of  simulated  problems  is  therefore  of  varying  complexity:  from
problems with a clear  structure when $\beta=0.45$ to problems without
any structure when $\beta=0.01$.  The experiments are performed twenty
times  for each value  of $\beta$  and the  average of  the normalized
mutual information  over these twenty simulated graphs  is depicted in
Figure  \ref{fig:simu1} for all  the algorithms together with the standard deviation using ribbons.  In  order to
produce  results as  comparable  as possible,  the  parameters of  the
different algorithms  were set  as follows: \textbf{vbmod,  mixer} and
\textbf{greedy ICL} were all started ten times and for each method the
best run  was selected according to the  corresponding model selection
criterion. The variational methods were  run with $K$ between 2 and 20
and the best clustering kept as a final result. For \textbf{greedy ICL}, the parameters of the prior $\eta^0,\zeta^0$ and $n_k^{0}$ were set to 1 and $K_{up}$ fixed to twenty. Finally the collapsed Gibbs sampler was run for 250 000 iterations (more than twice the default value).

The results illustrated in Figure \ref{fig:simu1} show that \textbf{greedy ICL}
outperforms the  other methods  for complex problems,  \emph{i.e.} low
values of $\beta$. The simulated clustering is recovered until $\beta$
reaches  $0.25$. Above  this  value the  different algorithms  perform
identically, but beyond this  limit the results of \textbf{greedy ICL}
are  a little bit  better. During  the transition  \textbf{greedy ICL}
gets slightly better results than the other algorithms, it is followed
by \textbf{colsbm}, \textbf{vbmod}, and the spectral baseline, which give close results.
\textbf{Mixer} deviates a little bit earlier from the planted clustering. On this simple setting all the algorithms are quite close.

\subsection{Setting 2: small scale community structures with a hub cluster}
\begin{figure}[h!]
\centering
\begin{tabular}{c}
\includegraphics[height=6.5cm]{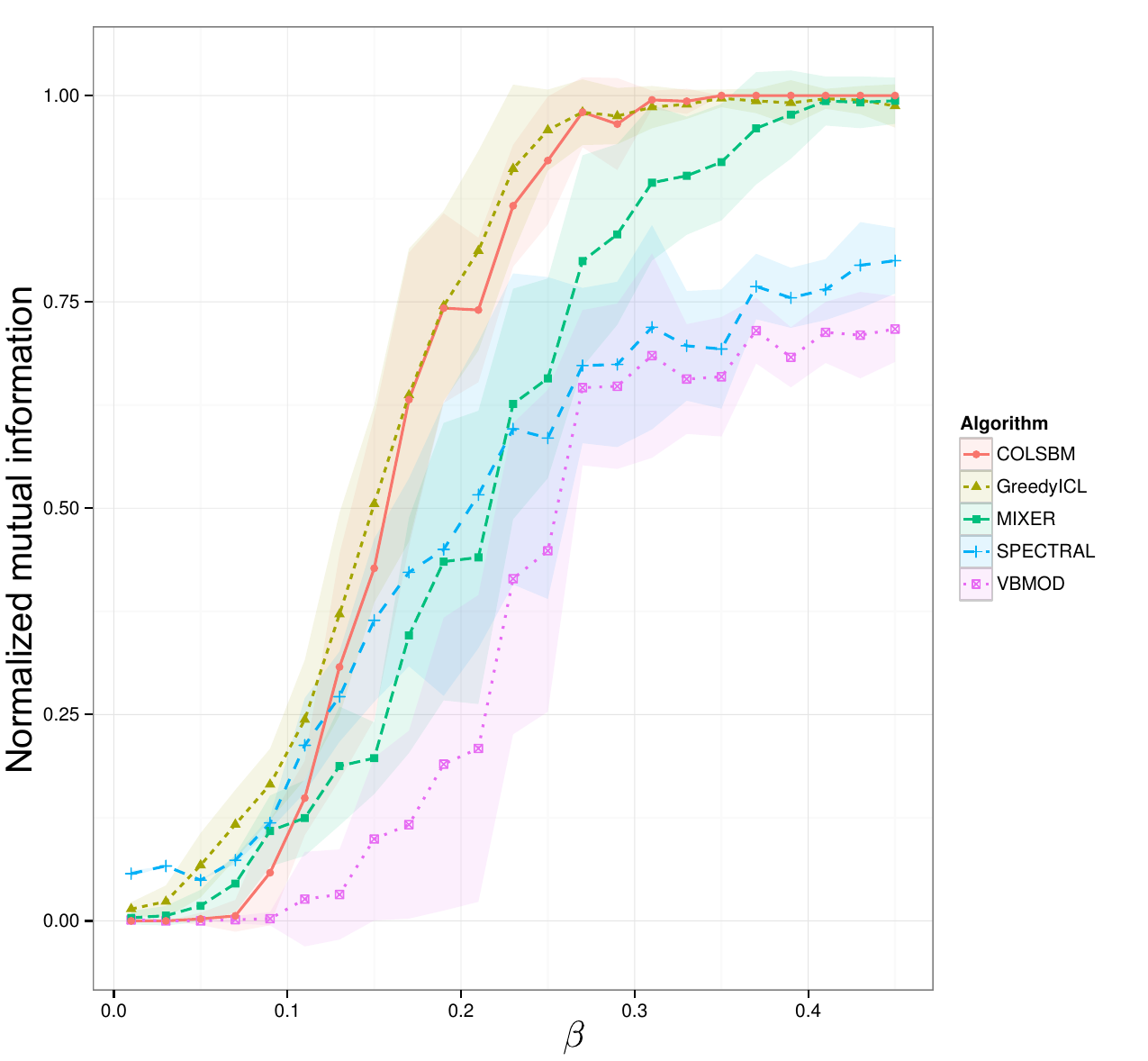}\\
 Setting 2
\end{tabular}
\caption{\label{fig:simu2}Mean of mutual information between estimated
  and true cluster membership matrices using $20$ simulated graphs for
  each value of  $\beta$ in $\{0.45,0.43,\hdots,0.03,0.01\}$, and with
  $N=100,   K=5,   \epsilon=0.01$   for   the   different   algorithms
  \textbf{greedy    ICL},    \textbf{vbmod},    \textbf{colsbm}    and
  \textbf{mixer}.  The spectral  clustering approach  was run  with the
  true number of clusters, and used as a baseline.}
\end{figure}
The second setting  aims at exploring the performances  of the methods
when the latent structure does not correspond only to communities. To this
end,  graphs were  generated  using the  stochastic  block model  with
affiliation probability matrix $\bPi$ of the form as in \cite{articlelatouche2012}:

$$\mathbf{\Pi}=\begin{pmatrix}\beta & \beta &\hdots &\hdots &\beta\\
\beta & \beta &\epsilon &\hdots &\epsilon\\
\beta & \epsilon &\beta  &\hdots &\epsilon\\
\beta & \epsilon & \hdots & \beta  &\epsilon\\
\beta & \epsilon &\hdots  &\hdots &\beta
\end{pmatrix}.$$

The clusters  correspond therefore to communities,  except one cluster
of  hubs  which  connects   with  probability  $\beta$  to  all  other
clusters.   Graphs   with  $N=100$   vertices,   $K=5$  clusters   and
$\balpha=(1/5,1/5,1/5,1/5,1/5)$ were  generated using this connection
pattern. The parameter $\epsilon$ was set to $0.01$ and $\beta$ ranged as previously from $0.45$ to $0.01$. Eventually, the other simulation parameters did not change. The results are shown in Figure \ref{fig:simu2}.

As  expected,  the  \textbf{vbmod}  algorithm, which  looks  only  for
communities,  is  strongly affected  by  this  change  of setting  and
systematically misses the hub  cluster. For the remaining methods, the
best results are achieved  by \textbf{greedy ICL} which still uncovers
the  planted  clustering  when  $\beta>0.25$,  whereas  \textbf{mixer}
starts  to drop  at $\beta$  equals 0.4.  The collapsed  Gibbs sampler
achieves also  good results  in this setting,  very close to  those of
\textbf{greedy  ICL} and  outperforms \textbf{mixer}.  Eventually, the
spectral  clustering  approach  suffers   from  the  same  problem  as
\textbf{vbmod}  and  tend  to  miss  the hub  class.  Note  that  for
difficult problems this method tends to perform slightly better. With respect to the variances of the results all the algorithms are quite comparable.

\subsection{Setting 3: medium scale community structures}
\begin{figure}[h!]
\centering
\includegraphics[height=6.5cm]{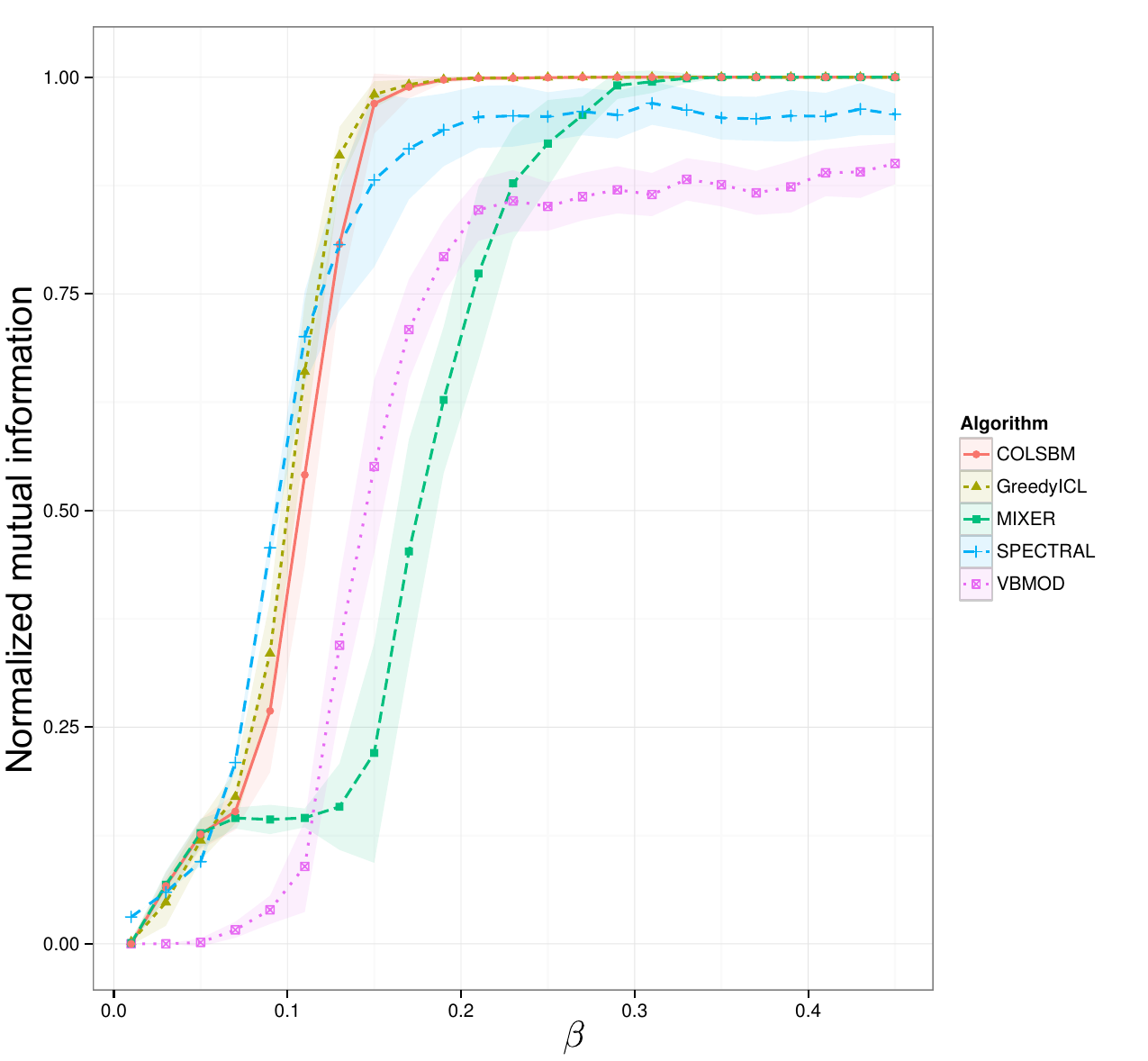}
\caption{\label{fig:simu2}
Mean of mutual information between estimated
  and true cluster membership matrices using $20$ simulated graphs for
  each value of  $\beta$ in $\{0.45,0.43,\hdots,0.03,0.01\}$, and with
  $N=500,   K=10,   \epsilon=0.01$   for  the   different   algorithms
  \textbf{greedy    ICL},    \textbf{vbmod},    \textbf{colsbm}    and
  \textbf{mixer}.  The spectral  clustering approach  was run  with the
  true number of clusters, and used as a baseline.}
\end{figure}

The third  setting is similar to Setting 2 (communities plus hub cluster) but  with more
nodes  and  clusters,  in order  to  study  the  effect of  these  two
parameters. Thus,  the number of vertices  was set to  $N=500$ and the
number of clusters to $K=10$.  The cluster proportions were defined as
$\balpha=(1/10,\hdots,1/10)$ and all the other parameters kept the
same  value  as  previously.  For this third experiment, the results presented in  Figure
\ref{fig:simu2} are very close between \textbf{greedy ICL} and \textbf{colsbm} which outperform the other approaches. \textbf{Mixer} also provide very good results until $\beta$ reaches 0.3 but then its results start to drop quite rapidly. While the spectral baseline and \textbf{vbmod} did not recover exactly the planted partition even when the problem was simple ($\beta$ around 0.4), they outperformed \textbf{mixer} when the planted structure was not particularly strong.    

The results obtained by the  different algorithms in this scenario  are better than
those  obtained  previously. The variances of the results are lower than in Setting 2. This  can  easily be  explained  by  the
increase in the number of nodes per cluster. The transitions between 
high and  low values  of the normalized  mutual information  were also
sharper than in the previous experiments, for the same reasons.

\subsection{Setting 4: large scale problem with complex structure}

\begin{figure}[h!]
\centering
\includegraphics[height=6.5cm]{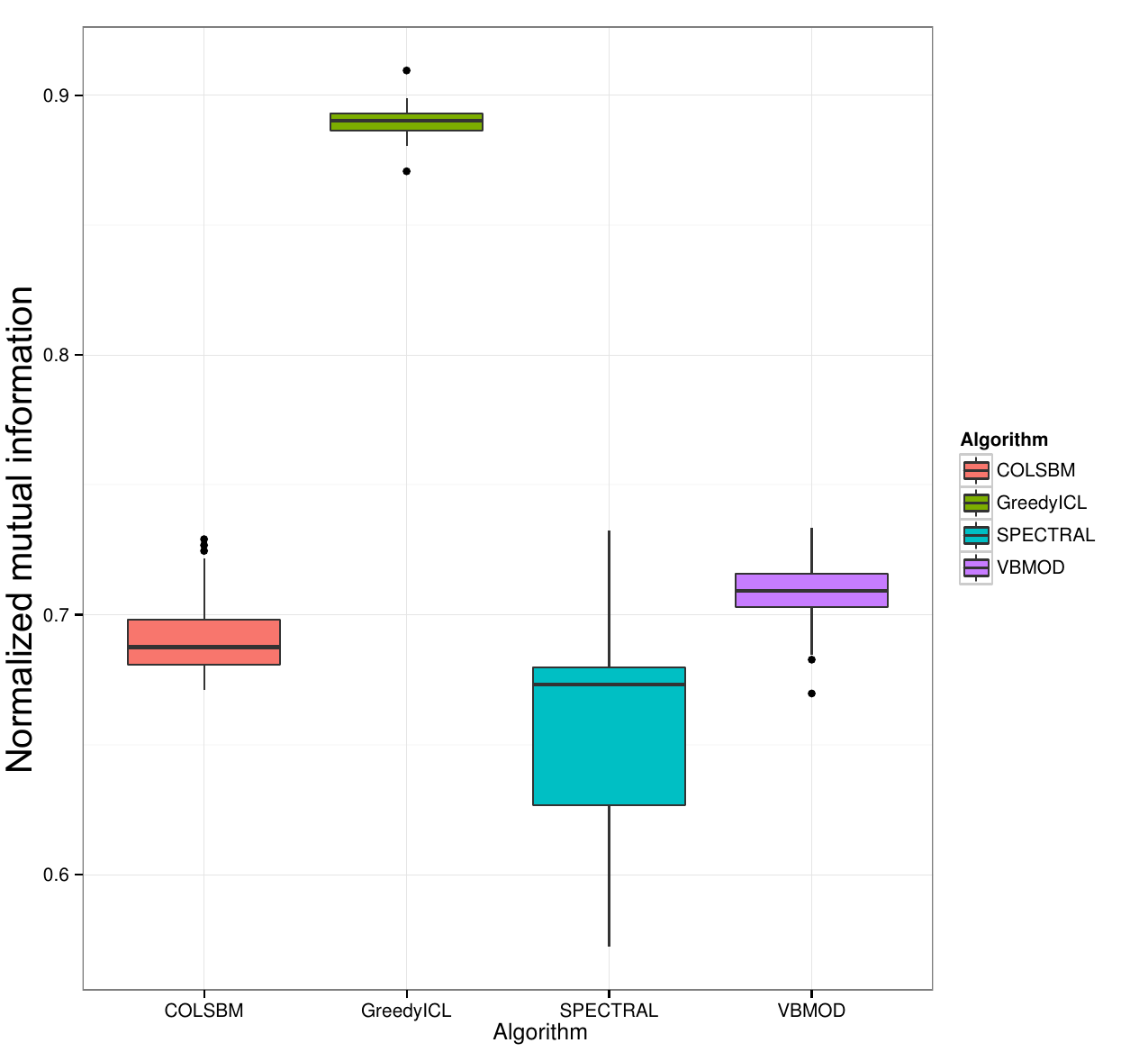}
\caption{\label{fig:simu4}Mean  of   the  mutual  information  between
  estimated and true cluster  membership matrices using $20$ simulated
  graphs with $N=10 000$ and $K=50$.  The spectral clustering approach
  along with \textbf{vbmod} were run with the
  true number of clusters, and used as a baseline.}
\end{figure}

The   final   setting   involves   larger  graphs   with   $N=10\,000$
vertices.  The  planted  structure  is  also not  a  purely  community
pattern.  Some interactions  between clusters  are  activated randomly
using a Bernoulli distribution as described by the following generative model: 
 
\begin{equation}
\Pi_{kl}=\begin{cases}ZU+(1-Z)\epsilon\textrm{, if $k\neq l$}\\
U\textrm{, if $k=l$}\\
\end{cases}
\end{equation}
with  $Z  \sim   \mathcal{B}(0.1)$,  $U  \sim  \mathcal{U}(0.45)$  and
$\epsilon=0.01$. The size of the problem and the complex nature of the
underlying structure, let only four algorithms able to deal with these
graphs namely \textbf{greedy ICL}, \textbf{colsbm}, \textbf{vbmod} and
spectral  clustering. \textbf{Mixer}  was not  tested since  it cannot
handle such large graphs. All approaches were used to cluster 20
simulated graphs generated using this scheme. The greedy algorithm was
started using $K_{up}=100$ and the  same setting as previously for the
prior  distributions.  The results  presented  as  boxplots in  Figure
\ref{fig:simu4} give a clear advantage to \textbf{greedy ICL} over all the remaining methods.
Thus,  \textbf{greedy  ICL}  achieves  an  average  normalized  mutual
information  of 0.88  whereas  \textbf{colsbm} reaches  only 0.67.  In
fact, the  greedy solution ended with  around 80 clusters  for all the
simulations whereas the Gibbs sampler  gives more than 240 clusters in
average and therefore produces highly over segmented partitions of the
graphs. Although they were supplied with the true number of clusters,
the  two  other approaches,  namely  \textbf{vbmod}  and the  spectral
method, give results clearly under those of \textbf{greedy ICL}, with an average normalized mutual information around 0.71 for the spectral method and 0.66 for \textbf{vbmod}.

To summarize the results we obtained in all the experiments we carried
out, it appears that  \textbf{greedy ICL} compares favourably with the
other existing  solutions for  SBM, in all  the settings.  The results
obtained in  complex setting, \emph{i.e.}  large graphs and  a complex
underlying  structure (Setting 4)  are particularly  encouraging since
\textbf{greedy  ICL} clearly outperforms  the collapsed  Gibbs sampler
and  the other  solutions even  if they  were provided  with  the true
number of cluster.

\section{Real dataset: communities of blogs}

The proposed algorithm was finally tested on a real network where vertices
 correspond to  blogs  and edges  to known  hyperlinks
between the blogs.  All the blogs considered are related to a common
topic, \emph{i.e.} illustrations and comics.

The  network   was  built  using  a   community  extraction  procedure
\citep{come10} which starts from known  seeds and expands them to find
a  dense core  of nodes  surrounding them.  It is  made of  1360 blogs
linked by 33 805 edges. The data set is expected to present specific
patterns, namely  communities, where two  blogs of the  same community
are more likely to be connected that nodes of different communities.
 To test this hypothesis we used the greedy ICL algorithm
and did a qualitative comparison of the results with those obtained with the 
 community discovery method of \cite{blondel08}.

\begin{figure}[h!]
\begin{center}
\begin{tabular}{c}
\includegraphics[width=10.5cm]{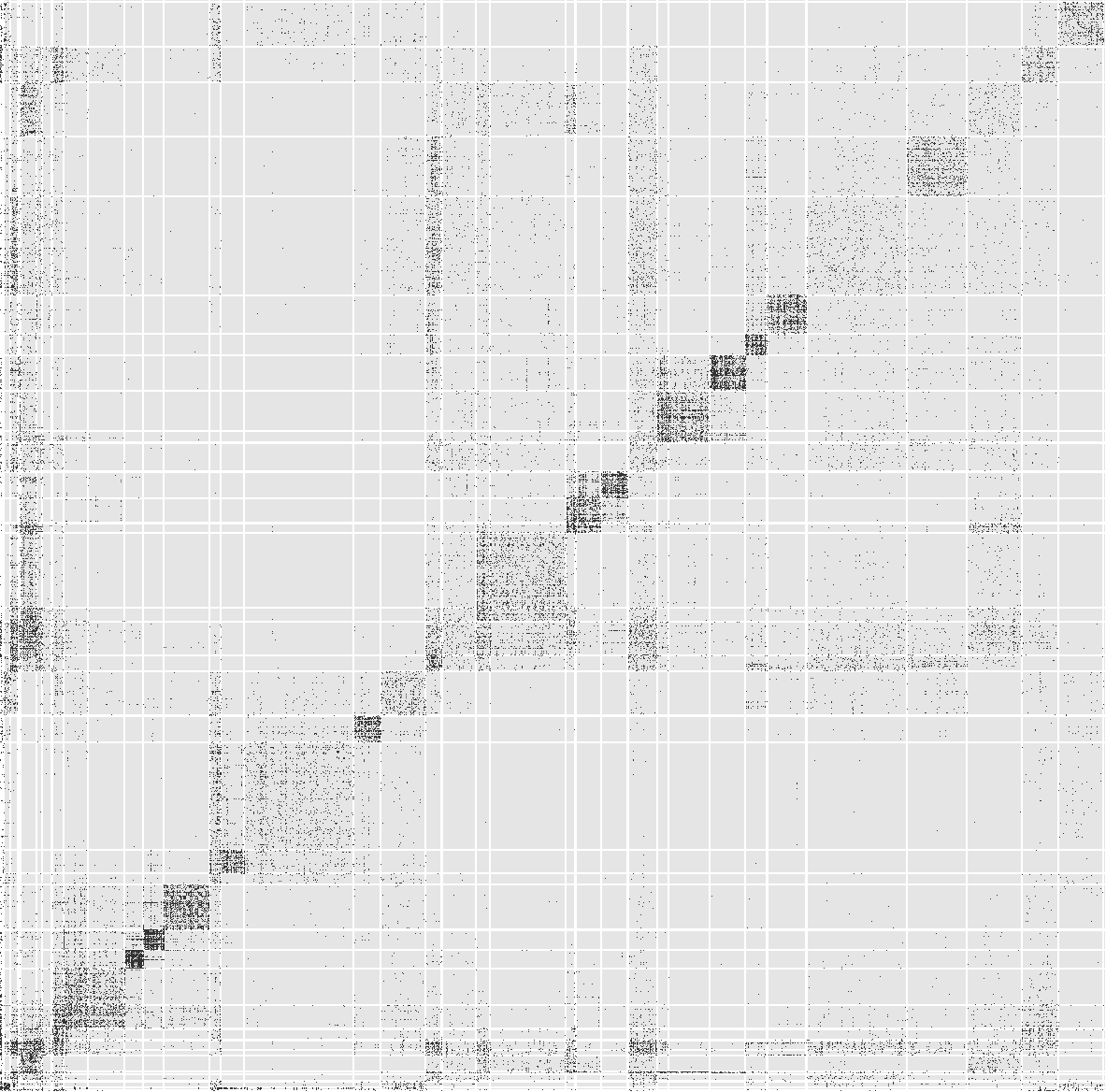}\\
\end{tabular}
\end{center}
\caption{\label{fig:datarealicl}Adjacency matrix of the network of blogs,
  the rows/columns are sorted by cluster number with clusters found by
  the greedy ICL algorithm. The cluster boundaries are depicted with white lines.}
\end{figure}

Starting   with  $K_{up}=100$  clusters,   greedy  ICL   found  $K=37$
clusters.  The  corresponding   clusters  are  illustrated  in  Figure
\ref{fig:datarealicl}  which is  an image  of the  adjacency matrix
with rows/columns sorted by cluster number.  Thus, it appears that the clusters found
correspond  in their  vast  majority to  small sub-communities.  These
sub-communities all  correspond to known groups. For  instance a group
of blogs of illustrators
for Disney was found. Other examples include clusters of blogs of
students who went to the same illustration school such as the ECMA school of
Angouleme or the ``Gobelins \'{E}cole de l'image''. However,
some clusters have more complex connectivity structures and are
made of hubs which highly connect to blogs of different clusters. They
correspond to blogs of  famous writers such as Boulet.

\begin{figure}[h!]
\begin{center}
\begin{tabular}{c}
\includegraphics[width=10.5cm]{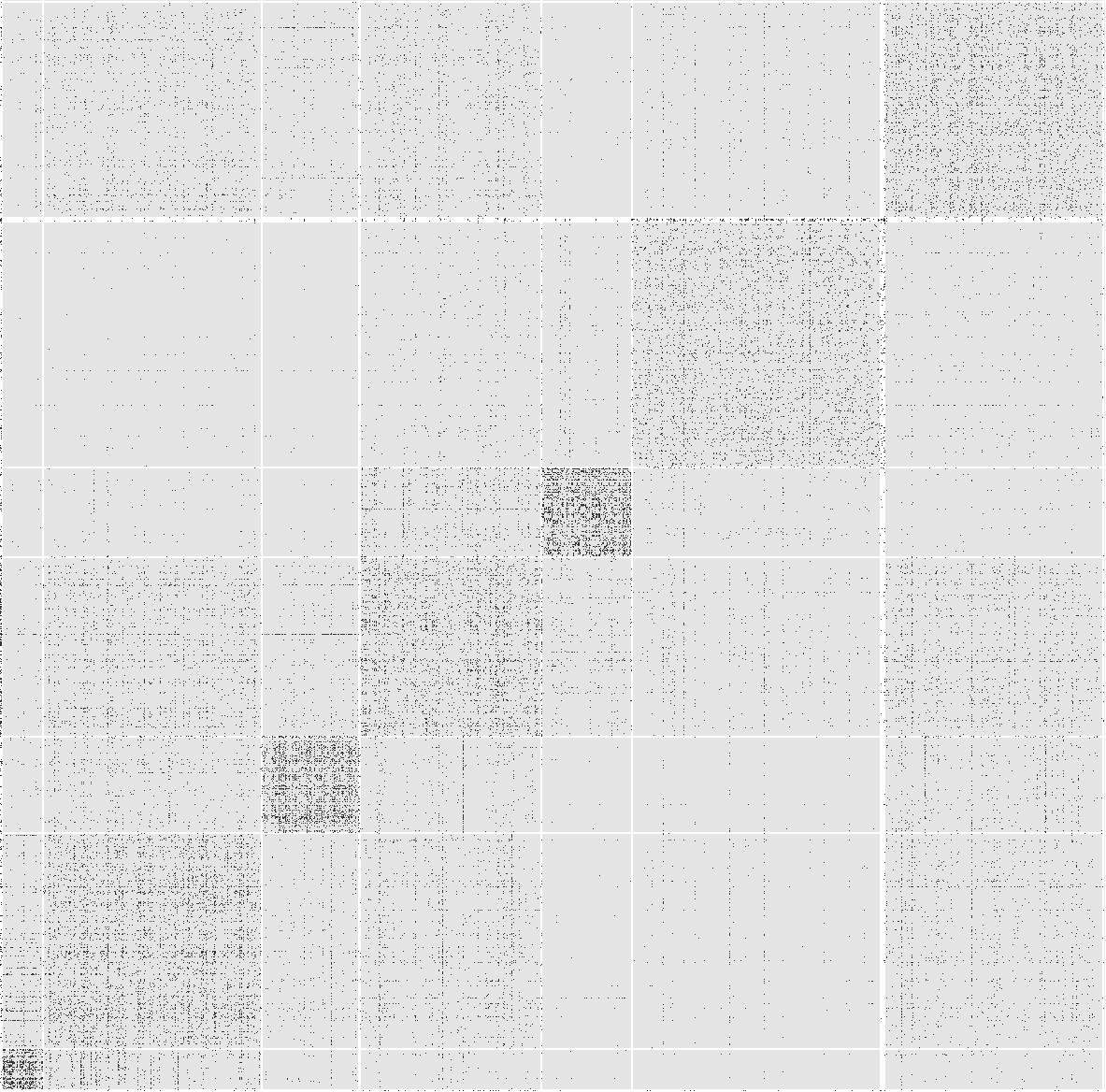}\\
\end{tabular}
\end{center}
\caption{\label{fig:datarealmodularity}Adjacency matrix of the network
  of blogs, the rows/columns are sorted by cluster number with clusters found by modularity optimization. The clusters boundaries are depicted with white lines.}\end{figure}

 To give a  qualitative idea of the interest  of the found clustering,
 we  also  give  the  results  obtained  by  the  community  discovery
 algorithm of \cite{blondel08} in Figure
 \ref{fig:datarealmodularity}. With this  approach only 8 clusters are
 found, corresponding  all to sub-communities. Clusters  of hubs could
 not be recovered. The major difference between the number of clusters
 estimated by the two methods  may be explained by two facts. Firstly,
 modularity  is  known to  be  prone  to  a resolution  limit  problem
 \citep{fortunato07} which  prevents such a solution  to extract small
 scale structures. This explains why the small sub-community extracted
 by \textbf{greedy ICL} are not recovered using the modularity. For the time being, the behaviour of the $ICL_{ex}$ criterion with respect to the resolution limit problem is not clear and will deserve further investigations. However, we notice that on this dataset finer structures than those obtained using modularity are recovered. Secondly, the difference in the way the two criteria use degree correction or not \citep{articlekarrer11} can also explain the disparity in the number of clusters. While modularity is a degree-corrected criterion which downscales the weights of the edges between highly connected vertices, the $ICL_{ex}$ criterion for the basic stochastic block model used here is not. Using a degree correction or not is a modelling choice which deserves to be validated and investigated; however, it seems that even without degree correction the results obtained by \textbf{greedy ICL} are meaningful, the hub clusters being interesting \textit{per se}.

\section{Conclusion}

In this paper, we relied on an analytical expression of the
integrated complete data log likelihood.  We then proposed a greedy optimization
algorithm  to  maximize this  exact  quantity.  Starting  from an  over
segmented partition, the approach simplifies the model, while
clustering the vertices, until a local
maximum is reached. This  greedy algorithm has a competitive complexity
and  may  handle networks  with  tens  of  thousands of  vertices  and
millions of edges.  We
illustrated on  simulated data that the method  improves over existing
graph clustering algorithms, both in terms
of  model selection  and clustering  of the  vertices.   A qualitative
comparison between methods was also carried out on an original network
 we built from blogs related to illustration, comics, and animations. 


\newpage
\appendix
\section*{Appendix}

\section{Integrated complete data log likelihood}
\label{ann:posterior}

Using  factorized and  conjugate  prior distributions  over the  model
parameters, the integrated complete data log likelihood is given by:
\begin{equation*}
\log p(\bX,\bZ|K) = \sum_{k,l}^{K} \log\left(
\frac{B(\eta_{kl},\zeta_{kl})}{B(\eta_{kl}^{0},\zeta_{kl}^{0})}\right) + \log\left(\frac{C(\bn)}{C(\bn^{0})}\right),
\end{equation*}
where 
\begin{itemize}
\item $\eta_{kl}=\eta_{kl}^{0}+\sum_{i \neq j}^{N} Z_{ik}Z_{jl}X_{ij}$
  for all $(k,l)$ in $\{1,\dots,K\}^{2}$
\item       $\zeta_{kl}=\zeta_{kl}^{0}       +      \sum_{i       \neq
    j}^{N}Z_{ik}Z_{jl}(1-X_{ij})$      for     all      $(k,l)$     in
  $\{1,\dots,K\}^{2}$
\item the components of the vector $\bn$ are $n_{k}=n_{k}^{0} +
\sum_{i=1}^{N}Z_{ik}$, for all $k$ in $\{1,\dots,K\}$
\item the
function           $B(a,b)$           is           such           that
$B(a,b)=\frac{\Gamma(a)\Gamma(b)}{\Gamma(a+b)}$  for  all  $(a,b)$  in
$\mathbb{R}^{2}$
\item the function
$C(\cdot)$                 is                 such                that
$C(\mathbf{x})=\frac{\prod_{k=1}^{K}\Gamma(x_{k})}{\Gamma(\sum_{k=1}^{K}x_{k})}$
for all $\bx$ in $\mathbb{R}^{K}$
\end{itemize}

{\bf Proof:}
Considering  factorized prior  distributions, the  integrated complete
data log likelihood decomposes into two terms:
\begin{equation}\label{eq:complete1}
  \begin{aligned}
\log p(\bX,\bZ|K)&=\log \left(\int_{\balpha,\bPi} p(\bX,\bZ,\bPi,\balpha|K)d\balpha
d\bPi\right) \\
&=                   \log\left(            \int_{\bPi}p(\bX|\bZ,\bPi,K)p(\bPi|K)d\bPi
\int_{\balpha}p(\bZ|\balpha,K)p(\balpha|K)d\balpha \right)\\
&= \log p(\bX|\bZ,K) + \log p(\bZ|K).
\end{aligned}
\end{equation}
The first term in (\ref{eq:complete1}) can be obtained from:
\begin{equation*}
\begin{aligned}
p(\bX|\bZ,K) &= \int_{\bPi}p(\bX|\bZ,\bPi,K)p(\bPi|K)d\bPi \\
&=  \int_{\bPi}\Big(\prod_{k,l}^{K}
      \Pi_{kl}^{\sum_{i      \neq      j}^{N}      Z_{ik}Z_{jl}X_{ij}}
      (1-\Pi_{kl})^{\sum_{i \neq j}^{N} Z_{ik}Z_{jl}(1-X_{ij})}\Big)
    \\
 &\times \prod_{k,l}^{K}\frac{1}{B(\eta_{kl}^{0},\zeta_{kl}^{0})}
    \Pi_{kl}^{\eta_{kl}^{0}-1}(1-\Pi_{kl})^{\zeta_{kl}^{0}-1}d\bPi \\
&=
\prod_{k,l}^{K}\left(\frac{B(\eta_{kl},\zeta_{kl})}{B(\eta_{kl}^{0},\zeta_{kl}^{0})}\int_{\Pi_{kl}}
  \mathrm{Beta}(\Pi_{kl};\eta_{kl},\zeta_{kl})d\Pi_{kl}\right) \\
&= \prod_{k,l}^{K}\frac{B(\eta_{kl},\zeta_{kl})}{B(\eta_{kl}^{0},\zeta_{kl}^{0})}.
\end{aligned}
\end{equation*}
The second term in (\ref{eq:complete1}) can be obtained from:
\begin{equation*}
  \begin{aligned}
    p(\bZ|K) &=  \int_{\balpha} p(\bZ| \balpha,K)p(\balpha|K) d\balpha
    \\
&=\int_{\balpha}\Big(\prod_{k=1}^{K}\alpha_{k}^{\sum_{i=1}^{N}Z_{ik}}\Big)\frac{1}{C(\bn^{0})}\prod_{k=1}^{K}\alpha_{k}^{n_{k}^{0}-1}d\balpha \\
    &=      \frac{C(\bn)}{C(\bn^0)}  \int_{\balpha}
    \mathrm{Dir}(\balpha;\bn) d\balpha \\
    &= \frac{C(\bn)}{C(\bn^0)}. 
  \end{aligned}
\end{equation*}
Finally,
\begin{equation*}
\log p(\bX,\bZ|K) = \sum_{k,l}^{K} \log\left(
\frac{B(\eta_{kl},\zeta_{kl})}{B(\eta_{kl}^{0},\zeta_{kl}^{0})}\right) + \log\left(\frac{C(\bn)}{C(\bn^{0})}\right).
\end{equation*}

\section{Change in ICL induced by a swap movement $i:g \rightarrow h$}
\label{ann:swap}
At  each step of the  greedy ICL  algorithm, a  single node  $i$ is
considered. If $i$ is currently in cluster $g$, the method tests every
possible label swapping $g\rightarrow h$, that is removing $i$ from cluster $g$ and
assigning it to a cluster $h \neq g$. The corresponding changes in the
$ICL_{ex}$ criterion are denoted $\Delta_{g \rightarrow h}$.  In order
to derive the calculation of each term $\Delta_{g \rightarrow h}$, for
all $h \neq  g$, we consider two cluster  indicator matrices $\bZ$ as
well as $\bZ^{test}$. $\bZ$ describes the current partition of
the  vertices  in  the  network,  while  $\bZ^{test}$  represents  the
partition after applying the swap $g\rightarrow h$:
\begin{equation*}
\left\{
\begin{array}{l}
  \bZ_{j}^{test} = \bZ_{j},\forall j \neq i \\
\\
 Z_{ik}^{test}=Z_{ik}=0,\forall k \neq g,h\\
\end{array}
\right.
\end{equation*}
while
\begin{equation*}
\left\{
 \begin{array}{l}
        Z_{ig}^{test} = 0, Z_{ig}=1 \\
\\
   Z_{ih}^{test} = 1, Z_{ih}=0
 \end{array}
\right.
\end{equation*}
Thus
\begin{equation*}
\Delta_{g \rightarrow h}=ICL_{ex}(\bZ^{test},K^{test})-ICL_{ex}(\bZ,K).
\end{equation*}
Note that  $\Delta_{g \rightarrow h}$ takes two  forms whether cluster
$g$ is empty after removing $i$ or not.
 In   the   later   scenario,   the   model   dimensionality   changes
 ($K^{test}=K-1$) and this must be taken into account to evaluate the possible increase induced by the swap movement.
\subsection{Case  1  : $\sum_{i}Z_{ig}^{test}>  0  $.  Cluster $g$  not
  empty after removing $i$}
\begin{eqnarray}
\Delta_{g \rightarrow h}&=&\log\left(\frac{C(\bn^{test})}{C(\bn)}\right)+\sum_{k,l}^{K}\log\left(\frac{B(\eta_{kl}^{test},\zeta_{kl}^{test})}{B(\eta_{kl},\zeta_{kl})}\right)\nonumber\\
&=&\log\left(\frac{\Gamma(n_g^{test})\Gamma(n_h^{test})}{\Gamma(n_g)\Gamma(n_h)}\right)+\sum_{l=1}^{K}\sum_{k\in
  \{g,h\}}\log\left(\frac{B(\eta_{kl}^{test},\zeta_{kl}^{test})}{B(\eta_{kl},\zeta_{kl})}\right)\nonumber\\
&+&\sum_{k\notin \{g,h\}} \sum_{l\in
  \{g,h\}}\log\left(\frac{B(\eta_{kl}^{test},\zeta_{kl}^{test})}{B(\eta_{kl},\zeta_{kl})}\right)\nonumber\\
&=&\log\left(\frac{\Gamma(n_g-1)\Gamma(n_h+1)}{\Gamma(n_g)\Gamma(n_h)}\right)+ \sum_{l=1}^{K}\sum_{k\in
  \{g,h\}}
     \log\left(\frac{B(\eta_{kl}+\delta_{kl}^{(i)},\zeta_{kl}+\rho_{kl}^{(i)})}{B(\eta_{kl},\zeta_{kl})}\right)\nonumber\\
&+&\sum_{k\notin \{g,h\}} \sum_{l\in
  \{g,h\}}\log\left(\frac{B(\eta_{kl}+\delta_{kl}^{(i)},\zeta_{kl}+\rho_{kl}^{(i)})}{B(\eta_{kl},\zeta_{kl})}\right)\nonumber\\
&=&\log\left(\frac{   n_h}{n_g-1}\right)+\sum_{l=1}^{K}\sum_{k\in
  \{g,h\}}\log\left(\frac{B(\eta_{kl}+\delta_{kl}^{(i)},\zeta_{kl}+\rho_{kl}^{(i)})}{B(\eta_{kl},\zeta_{kl})}\right)\nonumber\\
&+&\sum_{k\notin \{g,h\}} \sum_{l\in
  \{g,h\}}\log\left(\frac{B(\eta_{kl}+\delta_{kl}^{(i)},\zeta_{kl}+\rho_{kl}^{(i)})}{B(\eta_{kl},\zeta_{kl})}\right),\nonumber
\end{eqnarray}
with $\delta_{kl}^{(i)}$ the changes in edges counter $\eta_{kl}$ induced by the label swap:
\begin{eqnarray}
\delta_{kl}^{(i)}&=&\indi_{\{k=h\}}\sum_{j\neq i}^{N}Z_{jl}X_{ij}+\indi_{\{l=h\}}\sum_{j\neq i}^{N}Z_{jk}X_{ji}-\indi_{\{k=g\}}\sum_{j\neq i}^{N}Z_{jl}X_{ij}\nonumber\\&-&\indi_{\{l=g\}}\sum_{j\neq i}^{N}Z_{jk}X_{ji}.\nonumber
\end{eqnarray}
Moreover, $\rho_{kl}^{(i)}$ is defined in the following:
\begin{eqnarray}
\rho_{kl}^{(i)}&=&\left(\indi_{\{k=h\}} - \indi_{\{k=g\}}\right)(n_{l}-n_{l}^{0}-Z_{il})+\left(\indi_{\{l=h\}}-\indi_{\{l=g\}}\right)(n_{k}-n_{k}^{0}-Z_{ik})-\delta_{kl}^{(i)}.\nonumber
\end{eqnarray}
These update quantities can be  computed in $O(l_i)$ with $l_i$  the
degree of $i$ (total  number of edges from and to $i$). Therefore the average complexity of finding the best swap movement for a node is $\mathcal{O}(l+K^2)$, $l$ the average degree of the network for computing the $\delta_{kl}^{(i)}$ and $K^2$ to compute the $\Delta_{swap}$ with all the possible $h$ labels and keep the best one.  

\subsection{Case 2 : $\sum_{i}Z_{ig}^{test}= 0 $, cluster $g$ disappear}
In this case the dimensionality of $\bn^{0}$ changes and we will denote by $\bn^{0*}=(n^0,\hdots,n^0)$ the corresponding vector of size $K-1$.
\begin{eqnarray}
\Delta_{g \rightarrow h}&=&\log\left(\frac{C(\bn^{0})}{C(\bn)}\frac{C(\bn^{test})}{C(\bn^{0*})}\right)\nonumber\\
&+&\sum_{\stackrel{(k,l)\neq g}{k=h\:\mathrm{or}\:l=h}}\log\left(\frac{B(\eta_{kl}+\delta_{kl}^{(i)},\zeta_{kl}+\rho_{kl}^{(i)})}{B(\eta_{kl},\zeta_{kl})}\right)+\sum_{k
 =g\:\mathrm{or}\:l=g}\log\left(\frac{B(\eta_{kl}^{0},\zeta_{kl}^{0})}{B(\eta_{kl},\zeta_{kl})}\right)\nonumber\\
&=&\log\left(\frac{n_h}{n^{0}}\,\frac{\Gamma\left((K-1)n^0\right)\Gamma(Kn^0+N)}{\Gamma(K\,n^0)\Gamma((K-1)n^0+N)}\right) \nonumber\\
&+&\sum_{\stackrel{(k,l)\neq g}{k=h\:\mathrm{or}\:l=h}}\log\left(\frac{B(\eta_{kl}+\delta_{kl}^{(i)},\zeta_{kl}+\rho_{kl}^{(i)})}{B(\eta_{kl},\zeta_{kl})}\right)+\sum_{k
 =g\:\mathrm{or}\:l=g}\log\left(\frac{B(\eta_{kl}^{0},\zeta_{kl}^{0})}{B(\eta_{kl},\zeta_{kl})}\right).\nonumber
\end{eqnarray}
The complexity in this case is the same as previously \textit{i.e.} $\mathcal{O}(l+K^2)$.
\section{Change in ICL induced by a merge movement}
\label{ann:fusion}
\begin{eqnarray}
\Delta_{g \cup h}&=&\log\left(\frac{C(\bn^{0})}{C(\bn)}\frac{C(\bn^{test})}{C(\bn^{0*})}\right)\nonumber\\
&+&\sum_{\stackrel{(k,l)\neq g}{k=h\:\mathrm{or}\:l=h}}\log\left(\frac{B(\eta_{kl}+\delta_{kl}^{(i)},\zeta_{kl}+\rho_{kl}^{(i)})}{B(\eta_{kl},\zeta_{kl})}\right)+\sum_{k
 =g\:\mathrm{or}\:l=g}\log\left(\frac{B(\eta_{kl}^{0},\zeta_{kl}^{0})}{B(\eta_{kl},\zeta_{kl})}\right)\nonumber\\
&=&\log\left(\,\Gamma(n^{0})\frac{\Gamma\left((K-1)n^0\right)\Gamma(Kn^0+N)}{\Gamma(K\,n^0)\Gamma((K-1)n^0+N)}\frac{\Gamma(n_{h}+n_{g}-n^{0})}{\Gamma(n_{g})\Gamma(n_{h})}\right) \nonumber\\
&+&\sum_{\stackrel{(k,l)\neq g}{k=h\:\mathrm{or}\:l=h}}\log\left(\frac{B(\eta_{kl}+\delta_{kl}^{(i)},\zeta_{kl}+\rho_{kl}^{(i)})}{B(\eta_{kl},\zeta_{kl})}\right)+\sum_{k
 =g\:\mathrm{or}\:l=g}\log\left(\frac{B(\eta_{kl}^{0},\zeta_{kl}^{0})}{B(\eta_{kl},\zeta_{kl})}\right)\nonumber
\end{eqnarray}
with $\delta_{kl}^{(i)}$ the changes in edges counter $\eta_{kl}$ induced by the merge: 
\begin{eqnarray}
\delta_{kl}^{(i)}&=&\indi_{\{k=h\}}(\eta_{gl}-\eta_{gl}^{0})+\indi_{\{l=h\}}(\eta_{kg}-\eta_{kg}^{0})+\indi_{\{k=h\,\mathrm{and}\,l=h\}}(\eta_{gg}-\eta_{gg}^{0}).
\end{eqnarray}
Moreover, $\rho_{kl}^{(i)}$ is defined in the following:
\begin{eqnarray}
\rho_{kl}^{(i)}&=&\indi_{\{k=h\}}(\zeta_{gl}-\zeta_{gl}^{0})+\indi_{\{l=h\}}(\zeta_{kg}-\zeta_{kg}^{0})+\indi_{\{k=h\,\mathrm{and}\,l=h\}}(\zeta_{gg}-\zeta_{gg}^{0}).
\end{eqnarray}

\bibliography{biblio}

\end{document}